# On the Influence of Cognitive Styles on Users' Understanding of Explanations

*Short Paper*

**Lara Riefle**
Karlsruhe Institute of Technology
Karlsruhe, Germany
lara.riefle@kit.edu

**Patrick Hemmer**
Karlsruhe Institute of Technology
Karlsruhe, Germany
patrick.hemmer@kit.edu

**Carina Benz**
Karlsruhe Institute of Technology
Karlsruhe, Germany
carina.benz@kit.edu

**Michael Vössing**
Karlsruhe Institute of Technology
Karlsruhe, Germany
michael.voessing@kit.edu

**Jannik Pries**
Karlsruhe Institute of Technology
Karlsruhe, Germany
jannik.pries@alumni.kit.edu

## Abstract

*Artificial intelligence (AI) is becoming increasingly complex, making it difficult for users to understand how the AI has derived its prediction. Using explainable AI (XAI)-methods, researchers aim to explain AI decisions to users. So far, XAI-based explanations pursue a technology-focused approach—neglecting the influence of users' cognitive abilities and differences in information processing on the understanding of explanations. Hence, this study takes a human-centered perspective and incorporates insights from cognitive psychology. In particular, we draw on the psychological construct of cognitive styles that describe humans' characteristic modes of processing information. Applying a between-subject experiment design, we investigate how users' rational and intuitive cognitive styles affect their objective and subjective understanding of different types of explanations provided by an AI. Initial results indicate substantial differences in users' understanding depending on their cognitive style. We expect to contribute to a more nuanced view of the interrelation of human factors and XAI design.*

## Introduction

Over the last years, the constantly rising capabilities of artificial intelligence (AI) have paved the way to support users with increasingly complex tasks in a rising number of application domains (Kleinberg et al. 2018; Wang et al. 2019; Wu et al. 2019). For example, in medicine, AI has become capable of identifying certain clinical findings as accurately as human experts (Wu et al. 2019). However, as AI becomes more advanced, it also becomes more difficult to understand for users due to the highly complex nature of these algorithms (Zhang and Zhu 2018). This is especially problematic in high-stakes decision-making situations where the human user is held accountable for the final decision (Mohseni et al. 2021). Without understanding how the AI has derived its prediction, appropriately relying on its decisions becomes hardly possible (Mohseni et al. 2021).

To address this issue, research has worked on developing approaches to make an AI's decision more interpretable, which is reflected in the rising research field of explainable AI (XAI) (Adadi and Berrada 2018). The underlying idea of these approaches is to enable the user to better understand how an AI has derived its prediction. This should ideally result in an enhanced ability of the user to assess on a case-by-case basis when to rely on an AI's advice (Bansal et al. 2021; Zhang et al. 2020) as, on the one hand, it can help humans detect potential AI errors (Kenny et al. 2021). On the other hand, XAI can help humans calibrate trust toward the AI (Zhang et al. 2020). Despite the intended benefits of XAI, research has also revealed that explanations are often difficult to understand, which can thus lead humans to either follow incorrect AI advice or ignore correct suggestions (Bansal et al. 2021; Schemmer, Hemmer, Kühl, et al. 2022). A possible explanation for this finding might be grounded in the fact that explanation designers have predominantly focused on the algorithmic development of current XAI approaches without placing the end-users with their individual characteristics at the center of technology design (Ehsan, Wintersberger, et al. 2021). This means that the development of XAI approaches has so far generally followed a "one-size-fits-all" approach with all users being presented with the same explanations regardless of their individual way of thinking, perceiving, and remembering information, finally resulting in inferior decision-making performance. In this context, prior research demonstrated that when experienced and novice users receive the same explanations from the AI, this can lead to flawed decision outcomes (Szymanski et al. 2021). While experienced users could cope well with the explanations, novice users were not able to understand them and derived incorrect conclusions committing decision errors (Szymanski et al. 2021). Similarly, studies point out that different user groups prefer different presentation styles, e.g., visual or textual (Hernandez-Bocanegra and Ziegler 2021a; Szymanski et al. 2021).

Especially users' cognitive abilities and differences in information perception and processing are insufficiently addressed in XAI design (Schneider and Handali 2019). This observation is underpinned by Miller's (2019a) and Wang et al.'s (2019) seminal works, which call for incorporating research insights from cognitive psychology into XAI design. They propose that explanations should reflect human decision-making processes, i.e., they should consider users' limits of cognitive capacity and reasoning processes. While these studies remain conceptual, our research seeks to contribute empirical insights exploring the effect of users' cognitive abilities on their understanding of explanations. In particular, we draw upon the psychological construct of cognitive styles that describe humans' characteristic modes of processing information and approaching decision-making tasks (Hamilton et al. 2016; Kozhevnikov 2007). Research shows, for example, that humans either tend to solve tasks in a rational way, assessing all available information deliberately, or in an intuitive way, relying on their gut feelings (Hamilton et al. 2016; Kozhevnikov 2007). These fundamental human differences might impact users' approaches to understanding explanations. While some types of explanation might better match intuitive users' needs, others might be better suited for rational thinkers. Hence, we seek to answer the following research question:

*How do users' cognitive styles affect their understanding of different types of explanations provided by an AI?*

This paper presents the research design we apply to address this question as well as initial insights. We conduct a web-based experiment in which we present users with different types of explanations (i.e., example-based, rule-based, and feature-importance explanations) and measure their objective and subjective understanding. In addition, we assess users' cognitive styles, i.e., their characteristic mode of processing information, using questionnaires based on established psychometric scales. With our study, we expect to show the impact of users' individual differences in information processing on their ability to understand AI explanations correctly. By incorporating knowledge from cognitive psychology and providing initial empirical insights, we seek to contribute to the growing research field of XAI. Building on our results, researchers and practitioners will better understand the influence of individual characteristics on users' understanding of explanations and can thus adapt XAI system design accordingly.

## Background and Related Work

The advances in AI over the last years have paved this technology's entry into a continuously rising number of practical applications covering domains such as medicine (Wu et al. 2019), law (Kleinberg et al. 2018), or customer management (Leung et al. 2021). In many application domains, AI is employed to assist humans with the ultimate goal of improving the overall decision-making quality. Especially, its use in domains that involve decision-making with potentially high costs of errors, e.g., in medicine, has resulted in the requirement of increased transparency. That is, AI systems should not only support humans with recommendations for their decisions but also offer additional information allowing them to understand how a particular decision was derived by the AI (Rudin 2019). This requirement fueled the development of several methods to make AI's decisions more explainable resulting in the emergence of the research field of explainable AI (XAI) (Adadi and Berrada 2018). The underlying idea is that humans are enabled to assess whether to rely on the AI's prediction on a case-by-case basis.

By now, a broad range of **XAI approaches** has been developed, with feature importance methods (Ribeiro et al. 2016), example-based approaches (Cai et al. 2019), and rule-based explanations (Ribeiro and Guestrin 2018) being the most common ones. The idea of feature importance-based explanations is to quantify the contribution of each input variable to the prediction of a complex AI. Example-based explanations identify particular data instances of a dataset that are representative for the instance to be explained to provide users with guidance how a decision turned out for a comparable instance. Rule-based explanations methods aim to generate comprehensible descriptions of the AI's knowledge by deriving rules approximating its decision-making behavior (Adadi and Berrada 2018).

In accordance with the ongoing development of new XAI methods, research has also started to analyze whether evidence for the hoped-for utility of XAI approaches can be found using large-scale behavioral experiments (Schemmer, Hemmer, Nitsche, et al. 2022). In this context, several studies have analyzed the effect of additional information, e.g., explanations or AI confidence, on whether they can support humans to appropriately rely on the AI's predictions (Bansal et al. 2021). Whereas providing information about the confidence of an AI's decision has shown to be beneficial with regard to calibrating users' trust (Zhang et al. 2020), explanations can lead humans to either follow incorrect AI advice or ignore correct suggestions (Schemmer, Hemmer, Kühl, et al. 2022). This indicates that users might not have been able to establish a comprehensive understanding of the AI, which is key to effective usage and, in the end, improved decision-making performance (Bansal et al. 2021).

Possible reasons for why current XAI approaches have shown to provide only limited support for humans to rely on AI advice effectively might be found in the fact that the development of these algorithms has so far been mainly driven by an algorithmic perspective (Ehsan, Wintersberger, et al. 2021). Even though researchers have recently started to highlight the need for placing the human user at the center of explanation design (Ehsan, Passi, et al. 2021; Liao and Varshney 2021), the evaluation of XAI methods has so far predominantly pursued a technology-focused approach. However, this neglects the fact that humans' information processing is imperfect and that they have limited cognitive capacity, which needs to be considered when aiming at a human-centered XAI design (Miller 2019b).

The **influence of users and their characteristics on the human-system interaction** is one of the core IS research areas (Riefle and Benz 2021; Sidorova et al. 2008; Zhang and Li 2005). The frameworks by Zhang and Li (2005) and Rzepka and Berger (2018) structure the corresponding IS research, pointing out that not only the system but also the task, context and the human user influence the user interaction with AI. Consistent with task-technology-fit theory (Goodhue and Thompson 1995), they state that there must be a match of the functionality of the technology, the requirements of the task, and the characteristics and abilities of the individual (e.g., users' personality or cognitive abilities) to achieve the best possible interaction outcome (Goodhue and Thompson 1995; Rzepka and Berger 2018; Sidorova et al. 2008).

Individual user characteristics are stable over time and across task contexts (Kozhevnikov 2007) and can be defined as users' fundamental dispositions that determine how they perceive, think, feel, and behave (Kozhevnikov 2007; Zhang and Li 2005). Essentially, user characteristics can be divided into personality traits and cognitive abilities, which are crucial for humans' approach to problems and decision-making tasks (Kozhevnikov 2007; Zhang and Li 2005). Psychology describes and measures cognitive abilities with *cognitive styles*, defined as "consistencies in an individual's manner of cognitive functioning, particularly with respect to acquiring and processing information" (Kozhevnikov 2007, p. 464). At the onset of research on cognitive styles, researchers explored a range of cognitive style dimensions describing individual differences in perception (Kozhevnikov 2007). Studies in various contexts such as (managerial) decision-making (Sadler-Smith 2004), learning (Bostrom et al. 1990), or psychotherapy (Peterson et al. 1982) were conducted. As research progressed, a more integrated perspective on cognitive styles emerged (Kozhevnikov 2007). Thereby, research agrees upon two fundamentally different modes of processing information—namely, a rational and intuitive cognitive style (Hamilton et al. 2016; Novak and Hoffman 2009). An *intuitive* cognitive style is characterized by a contextualized and automatic decision-making process and reliance on first impressions and gut feelings (Hamilton et al. 2016). Intuitive thinkers tend to reason based on stereotypical thinking and associative connections, weighing experiences more than analysis (Hamilton et al. 2016). Studies find that intuitive thinking is schematic and heuristic and most effective when concrete examples are used (Novak and Hoffman 2009). In addition, intuitive thinkers often unconsciously react to salient features of decision situations (Hamilton et al. 2016). By contrast, a *rational* cognitive style is characterized by an analytical and deliberate decision-making process (Hamilton et al. 2016; Novak and Hoffman 2009). Rational thinkers thoroughly gather information, systematically evaluate the alternatives, and tend to reason based on logic (Hamilton et al. 2016). Prior research indicates that rational thinkers have a general motivation to engage in effortful cognitive activities (Hamilton et al. 2016).

The relevance of cognitive styles for user behavior in all kinds of contexts has been demonstrated in prior research across disciplines (Benbasat and Taylor 1978; Kozhevnikov 2007; Riefle et al. 2022). Research in psychology and education has shown that humans solve tasks differently depending on their cognitive abilities, which substantially impacts task performance, usage patterns, and user satisfaction. In IS research, cognitive styles have been found to influence the perception and evaluation of technological innovations (Chakraborty et al. 2008) or, in organizational contexts, the adoption of information systems and management performance (Benbasat and Taylor 1978). For example, Benbasat and Taylor (1978) noticed that analytical managers draw on different management reports for decision-making and tend to understand patterns in data better.

Nevertheless, research on cognitive styles in the context of XAI is scarce (Liao and Varshney 2021). First conceptual studies (Miller 2019b; Wang et al. 2019) point out that humans' use of explanations is impacted by their limits of cognitive capacity and decisions biases that can impair users' reasoning—which is why they call for incorporating knowledge about fundamental properties of human reasoning into XAI design. Some researchers have also started to empirically investigate the influence of users' cognitive abilities in the context of user-system interaction. Millecamp et al. (2020), for example, show that users' need for transparency and their interaction strategies with explanations of music recommendations differ depending on their rational or intuitive cognitive style. Summing up, cognitive styles are a crucial aspect to explore in the context of XAI. Only when we know the interrelations of users' rational or intuitive cognitive styles and their understanding of different explanation types can we enhance the understanding and lay the foundation for better performance outcomes of human-XAI system interaction.

## Research Model and Hypotheses

This research focuses on examining the influence of users' cognitive style on their understanding of different explanation types. Based on prior research in cognitive psychology, we develop hypotheses to investigate the interrelation of users' intuitive or rational thinking and their understanding of explanations. Figure 1 illustrates the included constructs and hypotheses, which will be described in the following.

The dependent variable in our research model is a user's objective and subjective understanding of different XAI-based explanation types. While subjective understanding comprises users' own evaluation of their understanding, objective understanding tests users' actual understanding by verifiable questions. Given that all explanation types have specific properties (Wang et al. 2019), we can assume that these properties will impact users' understanding. For example, prior IS research shows that different forms of information representation, e.g., textual or graphical, lead to different user perceptions and evaluation, e.g., in terms of effectiveness or liking (Hernandez-Bocanegra and Ziegler 2021b; Toker et al. 2013). Against this background, we hypothesize:

> H1: The type of explanation influences users' objective and subjective understanding of the explanation.

In addition, prior research in cognitive psychology suggests that people possess two fundamentally different modes of processing information and solving decision-making tasks—referred to as intuitive or rational cognitive style (Hamilton et al. 2016). In our case, these cognitive styles capture individual differences in the way users respond to presented explanations and how well they understand them. Hence, we can hypothesize that the effect of explanation types on users' understanding is moderated by users' rational or intuitive thinking. In particular, we hypothesize that certain properties of explanation types better match users' cognitive styles and thus enhance or diminish their understanding.

**Example-based explanations** explain an AI's prediction by selecting prototypical instances from the dataset (Adadi and Berrada 2018; Liao et al. 2020; Liao and Varshney 2021). These representative examples presented to the user are similar to the instance the user needs to make a prediction for and have the same predicted outcome or the alternative one (Liao et al. 2020; Liao and Varshney 2021). Given that intuitive thinkers tend to reason based on stereotypical thinking and associative connections, an intuitive cognitive style will likely increase users' understanding of example-based explanations. This is further substantiated by prior studies finding that intuitive thinking is schematic and heuristic and most effective when concrete examples are used (Novak and Hoffman 2009). While an intuitive cognitive style is characterized by weighing experiences and gut feeling more than analysis, the opposite holds for a rational thinking style. Rational thinkers tend to seek a lot of information and evaluate it systematically (Hamilton et al. 2016). As they rather avoid overgeneralization (Hamilton et al. 2016), example-based explanations will likely not match their cognitive style very well. In sum, we propose the following:

> H2a: Users' cognitive style moderates the effect of explanation type on understanding, such that *rational*/intuitive thinking *negatively*/positively influences the understanding of example-based explanations.

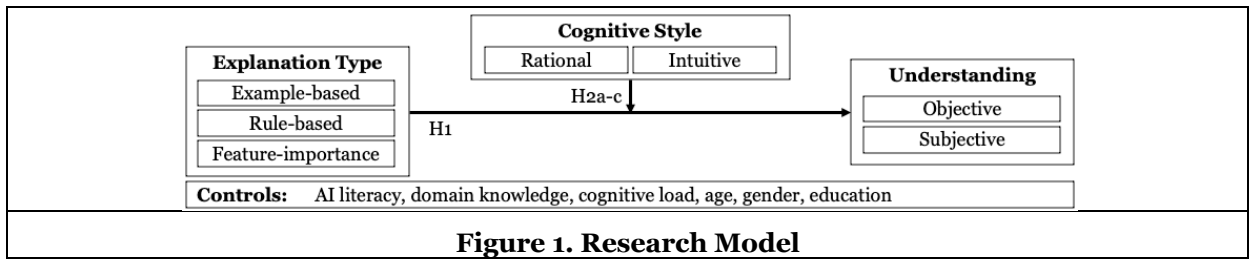

**Figure 1. Research Model**

In contrast to example-based explanations, **rule-based explanations** can be assumed to match rational thinkers' rule-governed reasoning (Hamilton et al. 2016; Novak and Hoffman 2009). Rule-based explanations approximate an AI's decision-making process by extracting logical clauses and combining them into rules (Adadi and Berrada 2018; Liao and Varshney 2021; Wang et al. 2019). These rules explain why the AI made its prediction based on the input features of the presented instance (Liao et al. 2020; Liao and Varshney 2021). When users execute the rules, this requires deductive reasoning (Wang et al. 2019), which fits well with a rational cognitive style (Novak and Hoffman 2009). Rational thinkers tend to think sequentially (Novak and Hoffman 2009) and make decisions in an intentional, reasoned way (Hamilton et al. 2016). A rational cognitive style is associated with the ability and preference to think in a logical way (Hamilton et al. 2016), which provides further substantiation to our hypothesis that rational thinking will positively influence the understanding of rule-based explanations. As outlined above, intuitive thinkers tend to rely on their initial hunches and do not enjoy logical reasoning (Hamilton et al. 2016). Hence, we hypothesize:

> H2b: Users' cognitive style moderates the effect of explanation type on understanding, such that *rational*/intuitive thinking *positively*/negatively influences the understanding of rule-based explanations.

**Feature-importance explanations** visualize how each feature of the presented instance contributes to the AI's prediction (Liao et al. 2020). In a multi-faceted representation of tabular data (i.e., the features) and bar graphs (i.e., the feature importance), this type of explanation provides different pieces of information to the user (Wang et al. 2019). Not only can users derive whether a feature had a positive or negative influence on the prediction, but they can also quantify the contribution of each feature (Adadi and Berrada 2018; Wang et al. 2019). These properties of feature-importance explanations likely resonate with a rational cognitive style. A rational cognitive style is characterized by analytical thinking and a preference and tendency for evaluating all available information thoroughly and systematically (Hamilton et al. 2016; Novak and Hoffman 2009). Moreover, rational thinkers enjoy cognitive challenges and prefer complex decision-making tasks over simple problems (Hamilton et al. 2016). Hence, rational thinkers will likely be able to cope well with the rather broad range of information provided by feature-importance explanations. Yet, as stated above, intuitive thinking associated with relying less on deliberate analysis in decision-making can be assumed to influence the understanding of feature-importance explanations negatively. Therefore, we formulate the following hypothesis:

H2c: Users' cognitive style moderates the effect of explanation type on understanding, such that *rational*/intuitive thinking *positively*/negatively influences the understanding of feature-importance explanations.

## Research Methodology

To explore how users' rational and intuitive cognitive style influence their understanding of different explanation types, we conduct a web-based experiment. By using the Internet for our research, we target a larger and more diverse sample, while internal validity is not negatively affected (Hergueux and Jacquemet 2015). An a priori power analysis using G*Power (Faul et al. 2007) with a significance level of 0.05 determined a minimum sample size of 206 participants to achieve a statistical power of 0.90 for detecting a medium effect size (f = 0.25).

### *Experiment Procedure & Measurement of Variables*

We test our research model by applying a between-subject design, in which each participant is randomly assigned to one of three conditions (i.e., example-based, rule-based, or feature-importance explanation). First, participants receive general instructions on the study procedure and are incentivized as described by Kvaløy et al. (2015). Next, they complete a training task—i.e., they are shown an exemplary instance of the dataset and explanation according to their treatment condition as well as exemplary questions that are later used to measure their objective understanding. In the main part of the study, participants are asked to complete four tasks. Similar to other XAI studies, we selected a churn prediction task and dataset from the domain of customer management (Leung et al. 2021). In each task, participants are presented with one instance from the dataset describing a customer, its characteristics, and the details of the current telco contract, as well as the AI's prediction (see Figure 2). Participants then need to decide whether the described customer will cancel the contract. Depending on the condition (see Figure 2), participants receive explanations that explain how the AI derived its prediction given the depicted features. Right after each decision task, users' objective understanding is measured. After completing four tasks, participants fill out a post-task questionnaire measuring their subjective understanding, followed by a final questionnaire measuring their cognitive styles, control variables, and demographics.

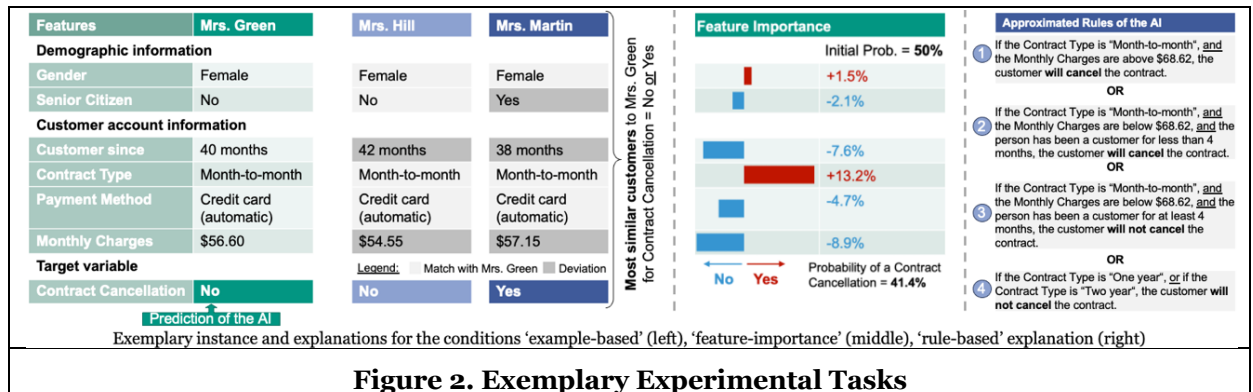

**Figure 2. Exemplary Experimental Tasks**

To measure the variables, we adapt established measures from prior literature. Objective understanding is measured using two questions based on Wang and Yin (2021) that assess participants' understanding in five facets: selecting the most/least influential features, specifying features' marginal effects on AI predictions, counterfactual thinking, simulating AI behavior, and detecting AI errors. For example, participants are asked, "If Mrs. Green would pay monthly charges of $75 and all other feature values remain the same, would this change the prediction of the AI?" and need to select the correct answer from the presented alternative options. To measure participants' subjective understanding, we adopt seven items based on Reijers and Mendling (2011) and Obar and Oeldforf-Hirsch (2020), e.g., "From the explanations, I understand how the AI works." Rational and intuitive cognitive styles are measured using the six-item scales by Hamilton et al. (2016). Finally, measures for the control variables are adapted from Klepsch et al. (2017) for cognitive load, from Ehsan et al. (2021) for AI literacy, and Hackbarth and Grover (2003) for domain knowledge. All items are measured on a 5-point Likert scale.

*Pre-Test and Initial Insights*

To validate our experiment design, we conducted a pre-test with 30 participants (average age 26.6; range: 18-40 years; 40% male; 57% holding at least a bachelor's degree; none stated to be an AI expert, 20% at least have some AI knowledge). Participants are recruited via the research platform prolific (www.prolific.co), as it provides access to a large and diverse participant pool and represents a trusted intermediary (Peer et al. 2017). An initial analysis of the data provides interesting insights into the relation of users' cognitive styles and their subjective and objective understanding of XAI-based explanation types (see Table 1): Depending on whether participants scored higher on the rational or intuitive cognitive style scale, they are classified as rational or intuitive thinkers. While rational thinkers (n=23) rated their *subjective* understanding of feature-importance explanations the highest (mean=3.98, SD=0.45), intuitive thinkers (n=7) rated rule-based explanations the highest (mean=4.14, SD=0.40). Yet, intuitive thinkers' *objective* understanding was lowest for rule-based explanations and feature-importance explanations (2 out of 4 correct) compared to example-based explanations. Focusing on users' *objective* understanding, both rational and intuitive thinkers scored the highest in the example-based explanation condition (2.75, respectively 2.50). However, users' perceived cognitive effort was also highest for example-based explanations—both for rational (mean=3.42, SD=0.62) and intuitive thinkers (mean=3.58, SD=0.35). Considering the control variable domain knowledge, 27% state to have no or almost no knowledge in the field of customer management, while 10% claim to be experts.

| Condition | Cognitive Style (n) | Subjective Understanding | | Objective Understanding* | | Cognitive Load | |
|---|---|---|---|---|---|---|---|
| | | Mean | SD | Mean | SD | Mean | SD |
| **Example-based** | Rational (8) | 3.30 | 0.65 | 2.75 | 0.46 | 3.42 | 0.62 |
| | Intuitive (2) | 2.57 | 0.61 | 2.50 | 0.71 | 3.58 | 0.35 |
| **Rule-based** | Rational (8) | 3.82 | 0.36 | 1.88 | 0.64 | 2.96 | 0.35 |
| | Intuitive (2) | 4.14 | 0.40 | 2.00 | 1.41 | 2.67 | 0.24 |
| **Feature-importance** | Rational (7) | 3.98 | 0.45 | 2.00 | 1.15 | 2.98 | 0.15 |
| | Intuitive (3) | 3.71 | 1.22 | 2.00 | 1.00 | 3.00 | 0.44 |
| * Measured by number of correct answers out of 4 possible. | | | | | | Marks the highest values per variable. | |

**Table 1. Initial Results**

## Conclusion and Outlook

In this paper, we argue for investigating users' objective and subjective understanding of different XAI-based explanations depending on their rational and intuitive cognitive styles. By conducting a web-based behavioral experiment, we aim to contribute to a more nuanced view of how users' characteristic modes of information processing interrelate with XAI design. So far, we have conducted a pre-test of the experiment, in which participants are randomly presented one of the most commonly employed XAI-based explanation types, i.e., example-based, rule-based, and feature-importance explanations, and both their objective and subjective understanding of these are measured. Initial results indicate substantial differences in users' understanding depending on their cognitive style. Next, we will collect data from a larger sample to statistically test our hypotheses. Thereby, we expect to advance research in the growing field of XAI. Given the prevailing assumption that explanations are understood equally by all users, with our research, we aim to provide empirical insights to support the call for a more human-centered XAI design (Miller 2019b; Wang et al. 2019). Building on our findings, XAI designers will be enabled to enhance explanations both on an algorithmic and information presentation level to foster users' understanding of AI's predictions.